\documentclass[10pt,twocolumn,letterpaper]{article}

\usepackage{cvpr}             

\usepackage{multirow}
\usepackage{mathtools}
\usepackage{amsmath}
\usepackage{soul}
\usepackage{graphicx}
\usepackage{xcolor}  
\usepackage{overpic}
\usepackage{wrapfig}
\usepackage{siunitx}
\usepackage[accsupp]{axessibility}

\DeclareMathOperator*{\argmin}{arg\,min}
\DeclareMathOperator*{\tr}{tr}
\DeclareMathOperator*{\Hess}{Hess}
\DeclareMathOperator*{\Vector}{vec}
\DeclareMathOperator*{\grad}{grad}

\newtheorem{prop}{Proposition}

\definecolor{cvprblue}{rgb}{0.21,0.49,0.74}
\usepackage[pagebackref,breaklinks,colorlinks,allcolors=cvprblue]{hyperref}

\def\paperID{***} 
\def\confName{CVPR}
\def\confYear{2026}

\title{FreeForm: Reduced-Order Deformable Simulation\\from Particle-Based Skinning Eigenmodes}

\author{
Donglai Xiang$^{1,*}$ \qquad
Vismay Modi$^{1,*}$ \qquad
Rishit Dagli$^{1,2,\dagger}$ \qquad
Ty Trusty$^{1,2,\dagger}$ \qquad 
Gilles Daviet$^{1}$ \qquad \\
Anka He Chen$^{1}$ \qquad
Nicholas Sharp$^{1}$ \qquad
David I.W. Levin$^{1,2}$ \vspace{2mm}\\
$^{1}$NVIDIA \qquad
$^{2}$University of Toronto \vspace{1mm}\\
{\footnotesize $*$ Equal contribution \qquad $\dagger$ Work done during internship at NVIDIA}
}

\begin{document}


\twocolumn[{%
\renewcommand\twocolumn[1][]{#1}%
\maketitle
\begin{center}
    \centering
    \captionsetup{type=figure} 
    \includegraphics[width=\textwidth]{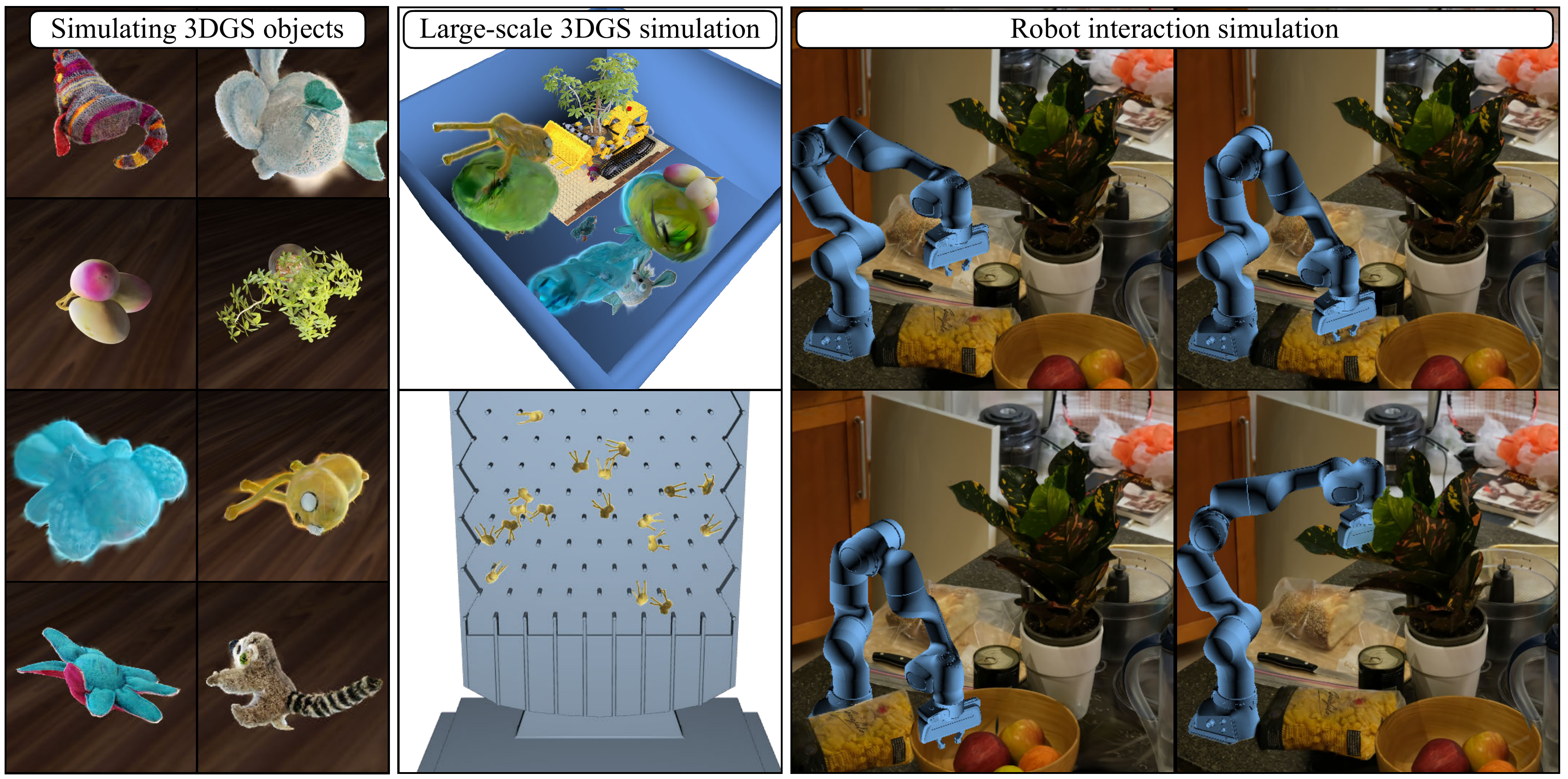}
    \captionof{figure}{Left: we show results of our reduced-order elastic simulation applied to 3D Gaussian Splatting (3DGS) objects. Middle: our simulation can handle multiple interacting 3DGS objects. Right: we show the application of our method in simulating robot interaction.}
    \label{fig:teaser}
\end{center}%
}]

\begin{abstract}
We present a novel formulation for mesh-free, reduced-order simulation of deformable hyperelastic objects. Existing work in reduced-order elastodynamic simulation represents the input geometry by either meshes, which can be difficult to obtain due to challenges in scanning and triangulating complex shapes, or by neural fields that require per-shape optimization.
We propose to adopt a Reproducing Kernel Particle Method (RKPM) representation, which enables the construction of reduced-order skinning weights by solving a generalized eigensystem on the Hessian matrix of the elastic energy. We demonstrate that this formulation not only leads to a 40$\times$ training speedup compared with the per-shape optimization of neural fields, but also achieves lower simulation error when evaluated against the converged results of finite element method. We show our simulation results on a wide variety of objects in different representations including meshes and Gaussian splats, as well as the application of our method in the downstream task of robot simulation.
\end{abstract}
    
\section{Introduction}
\label{sec:intro}

Elastodynamic simulation of deformable objects is important and widely used in engineering, scientific computing, visual effects, and robotics.
The Finite Element Method~(FEM) is typically employed to this effect; however it suffers from two major limitations. First, element-based FEM requires high-quality meshes as input: this can be problematic on traditional mesh representations due to the challenges of volumetric meshing on arbitrary shapes, and may not even be well-defined for modern, imprecise point-based representations such as Gaussian Splats~\cite{kerbl3Dgaussians}. 
Second, an accurate FEM simulation at high resolution needs a similarly high number of Degrees of Freedom~(DoFs) and more iterations of numerical solves, and is thus slow to compute. 

Particle-based methods like the Material Point Method~(MPM) and Smoothed Particle Hydrodynamics~(SPH) have been proposed, which can simulate elastic objects in a mesh-free manner, and are popular in recent works that aims to simulate objects represented by 3DGS, e.g, PhysGaussian and PhysDreamer~\cite{xie2023physgaussian,zhang2024physdreamer}. However, these approaches also have limitations; for example, they are sensitive to both spatial and temporal discretizations~\citep{yue2015continuum,desbrun1999spacetime}, potentially leading to failures under large strains.

Meanwhile, reduced-order simulation techniques \cite{benchekroun2023fast} have been proposed to reduce computation costs, but have mainly been focused on mesh-based simulation. A recent work, Simplicits \cite{modi2024simplicits}, addresses the problem of reduced-order simulation in the mesh-free domain, the same setting as our work. This approach, however, requires optimizing a neural field for every input object before simulation can be run. Moreover, we empirically find that Simplicits achieves suboptimal simulation accuracy, possibly due to the difficulty in variational optimization of elastic energy (see discussion in the experiments).

To address the limitation of previous work, our key insight is that the Reproducing Kernel Particle Method (RKPM)~\cite{liu1995reproducing}, a mesh-free, particle-based representation of spatially-defined functions, enjoys multiple advantages in this setting, to which it has not previously been applied to the best of our knowledge. We leverage RKPM to parameterize the deformation subspace and formulate the elastic energy in a mesh-free manner. More importantly, this explicit representation makes it possible to obtain a set of optimal skinning eigenmodes through eigenanalysis, which is more accurate and significantly faster than other comparable subspace generation techniques.

Our contributions are summarized as follows:

\begin{itemize}
    \item We present a novel formulation of mesh-free, reduced-order elastodynamics using skinning eigenmodes of elastic objects parametrized by RKPM;
    \item We derive a simple, easy-to-implement mathematical expression for the Hessian matrix of the commonly used Neo-Hookean elastic energy. 
    \item We empirically demonstrate the efficiency and effectiveness of our method on a wide variety of objects.
\end{itemize}

\section{Related Work}
\label{sec:related}
Traditionally, in graphics and engineering, elastodynamic simulations have employed explicit mesh-based representations of objects, optimizing over per-element energies ~\cite{terzopoulos1987elastically,macklin2016xpbd,bouaziz2014projective}. However, the recent popularity of NeRFs~\cite{Mildenhall20eccv_nerf}, Gaussian Splats~\cite{kerbl3Dgaussians}, and signed-distance functions have given rise to new simulation techniques naturally supporting these implicit object representations.

\paragraph{Neural physics-based simulation}
Neural physics simulation is typically motivated by the need to simulate geometries or materials for which traditional pipelines fail or for which parameters or specifications are not known~\citep{xie2023physgaussian,jiang2024vr-gs,rong2024gaussiangarments,romero2021subspacelearning,whitney2023learning3dparticlebasedsimulators}. These methods exist on a spectrum, ranging from augmenting existing simulation algorithms with neural components~\citep{romero2022contactcentric,daviet25neuralfem,holden2019subspace} to replacing physics simulation in its entirety with a learned representation~\citep{li2018learning,pfaff2021learning}. Such algorithms are of increasing usefulness due to the proliferation of new, high-fidelity geometric data representations~\cite{Mildenhall20eccv_nerf,kerbl3Dgaussians} and a desire to generate physically plausible motion from them, whether that be for games~\cite{jiang2024vr-gs}, training other neural models~\cite{ling2024align} or applications in the physical AI~\cite{abou-chakra2024physically, xu2025neural}.

\begin{figure*}[t]
    \centering
    \includegraphics[width=0.9\textwidth]{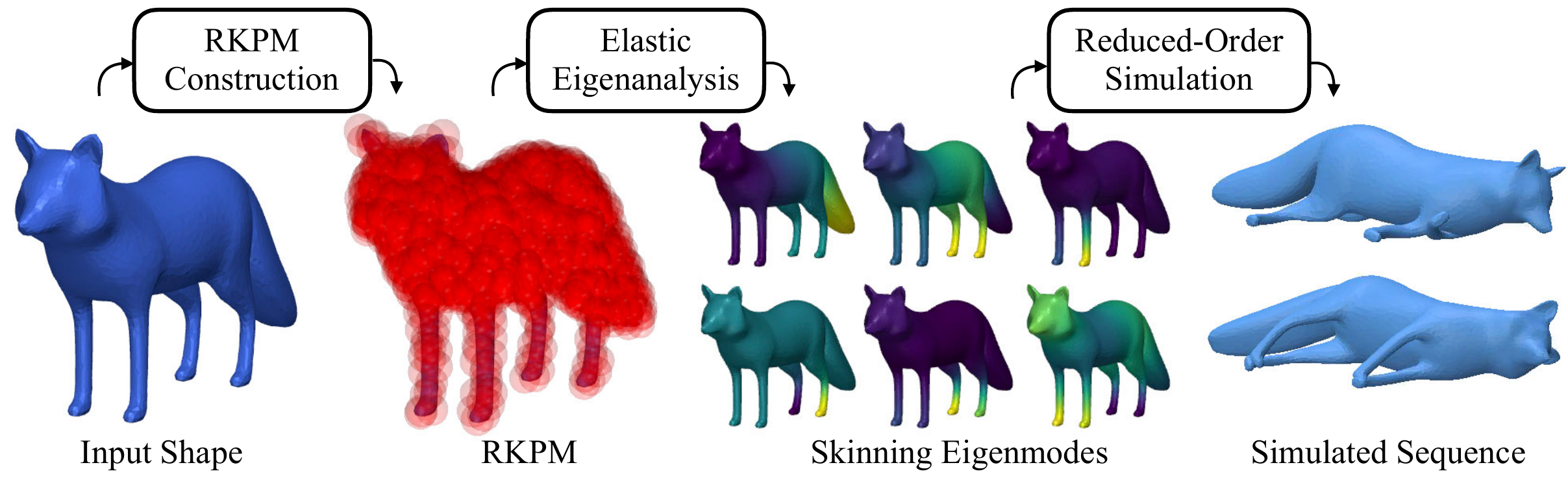}
    \caption{Overview of our method. Given an input object’s shape and material properties, we first construct RKPM particles and perform eigenanalysis to derive expressive skinning weights. These weights are then used to enable reduced-order simulation at runtime.}
    \label{fig:method_overview}
\end{figure*}

\paragraph{Particle-based simulation.}
Alternatively, particle-based updated-Lagrangian methods, such as the Material Point Method~\cite{SulskySolidPIC,Jiang2016mpm} and Smoothed Particle Hydrodynamics~\cite{gingold77sph,Desbrun1996} have been employed to simulate Gaussian Splats or NeRFs~\citep{xie2023physgaussian, Li2023PACNeRFPA}. These methods can handle a wide range of constitutive models. However, their sensitivity to the spatial and temporal discretizations renders them non-ideal for elastic solid simulation, with numerical fracturing under large strains and inexact boundaries as common limitations~\citep{Jiang2016mpm}; RKPM and Moving Least Squares (MLS) interpolation techniques have been explored to reduce these artifacts~\citep{westhofen23sph, chen2020mlsrkpm}. \citet{zong2023neural} explored reducing the computational cost using neural fields.
\citet{martin2010elastons} proposed a fully implicit particle-based simulation method leveraging generalized MLS; however, this suffers from a high DoF count and lack of convergence with particle resolution. \citet{feng2023pienerf} reduce the DoF count using clustering, but lacks awareness of geometric details.
Position-Based Dynamics and mass-spring techniques have also been employed to simulate Gaussian particles~\citep{jiang2024vr-gs,abou-chakra2024physically,zhong2024springgaus}, but do not derive from a proper continuum model and as such require the use of ad-hoc material parameters.
\citet{dodik2024robust} introduce a particle-based biharmonic formulation with motivations similar to ours, but focus on skinning control from predefined handles rather than physical simulation.

\paragraph{Reduced-order simulation.}
Rather than rely on a large number of particles or nodes to represent motion, reduced approaches rely on a small set of degrees of freedom augmented with complex basis~\cite{barbi2005subspace} or interpolating functions~\citep{Wang2015}. These methods can produce rich shape aware motion as long as appropriate basis functions can be computed, via modal analysis~\citep{BEH18,benchekroun2023fast, yang2015expediting} or exemplars~\citep{barbi2005subspace}.
Training a neural network to encode the reduced representation has shown great promise for modeling kinematics and dynamics of complex geometries~\citep{Fulton2019,sharp2023data,lyu2024accelerate}. While LiCROM~\citep{chang2023licrom} learns a continuous reduced order model from an existing subspace --- usually generated on a mesh, Simplicits~\citep{modi2024simplicits} demonstrates representation agnostic simulation across any input domain that admits an inside-outside query, along with shape-aware deformation and good agreement with standard FEM simulations. \citet{chang2025shape} trains a neural field that predicts Laplace eigenfunctions for classes of parametric objects, which may also be used for reduced order simulation. Besides the requirement of parametric models, unlike Simplicits~\citep{modi2024simplicits} and our method, this approach is not material-aware and thus only addresses homogeneous objects.

\section{Methodology}

Our method takes as input the geometry of an elastic object, in any representation that allows us to do integration on the volume, e.g, by sampling points from the object.
First, in an optimization/training stage, we construct a set of reduced-order bases, known as \emph{skinning weights}, for the object.
Then, in a simulation stage, the skinning weights are used for a low-degree-of-freedom elastic simulation. 
We introduce the overall background and notation in Sec.~\ref{sec:method_background}, and our approach in Sec.~\ref{sec:method_rkpm}. 
A high-level overview of the method is given in Fig.~\ref{fig:method_overview}.

\subsection{Background}
\label{sec:method_background}

Reduced-order elastodynamic simulation models the deformation of an object by a deformation map $\mathbf x \gets \phi(\mathbf X, \mathbf z)$ that maps any point $\mathbf X \in \Omega \subset \mathbb{R}^3$ in the object from the reference space to $\mathbf x \in \mathbb{R}^3$ in the deformed configuration, controlled by a number of Degrees of Freedom (DoFs) $\mathbf z \in \mathbb{R}^n$.
In a maximal-coordinate simulation, the DoFs are simply the vertex or particle positions, while reduced-order simulations pick DoFs from a low-dimensional space for efficiency and control.
One common formulation is skinning-based (or frame-based) deformation \cite{gilles2011frame, benchekroun2023fast, dodik2024robust}, where the DoFs $\mathbf z$ consist of a set of $m$ affine transformations $\{\mathbf Z_j \in \mathbb{R}^{3 \times 4}\}_{j=1}^m$ with associated fixed skinning weight functions $\mathbf W: \mathbb{R}^3 \rightarrow \mathbb{R}^m$ combined using Linear Blend Skinning (LBS):
\begin{gather}
\mathbf x = \Phi(\mathbf X, \mathbf z) = \mathbf X + \sum_{j=1}^{m} \mathbf W^j(\mathbf X) \mathbf Z_j \overline{\mathbf X},
\label{eq:method_skinning}
\end{gather}
where $\overline{\mathbf X}$ is the homogeneous coordinates of $\mathbf X$, $ \mathbf W^j$ is the $j$-th weight of $\mathbf W$. 

Reduced-order simulation methods have a \textit{fitting} or \textit{training stage} that finds skinning weights $\mathbf W$ for a given object, and a \textit{simulation stage} that, given $\mathbf W$, the previous state $\mathbf{z}_{t+1}$, and the environment, performs time stepping to solve for DoFs of the next state $\mathbf z_{t+1}$.
The simulation stage follows the standard implicit time integration that minimizes the incremental potential:
\begin{gather}
\mathbf z_{t+1} = \arg \min_\mathbf z \text{Ir}(\mathbf z, \mathbf z_t) + E_{\text{pot}}(\mathbf z) + E_{\text{ext}}(\mathbf z)
\label{eq:simplicit_sim}
\end{gather}
where $h$ is the timestep, $\text{Ir}(\mathbf z, \mathbf z_t)$ is the inertia energy, $E_{\text{pot}}$ is the elastic potential energy, and $E_{\text{ext}}$ is the potential energy for external forces such as gravity and boundary conditions, both implicitly dependent on the skinning model $\mathbf W$.


\paragraph{Simplicits}
One baseline, Simplicits, proposes a mesh-free formulation by representing the skinning weights $\mathbf W$ with a neural field parametrized by network weights $\theta$ (hence the notation $\mathbf W_\theta, \Phi_\theta$). Simplicits aims to find the skinning weights that produce physically-plausible deformation by minimizing a combination of elastic loss and orthogonality constraints:
\begin{gather}
\theta^* = \arg \min_\theta \lambda_{\text{elastic}} \mathcal{L}_{\text{elastic}} + \lambda_{\text{ortho}} \mathcal{L}_{\text{ortho}}
\label{eq:simplicit_loss_sum}
\end{gather}
The elastic loss evaluates the elastic energy of the deformation map $\Phi_\theta$ over randomly sampled transformations $\mathbf z$ from a normal distribution of variance $\sigma^2$
\begin{equation}
\begin{gathered}
\mathcal{L}_{\text{elastic}} = \mathbb{E}_{\mathbf z \sim \mathcal{N}(\mathbf 0, \mathbf \sigma \mathbf I)}\left[ E_{\text{pot}}(\mathbf z) \right] \\
= \mathbb{E}_{\mathbf z \sim \mathcal{N}(\mathbf 0, \sigma \mathbf I)}\left[ \int_{\Omega} \Psi(\Phi_\theta(\mathbf X, \mathbf z)) d\mathbf X \right],
\label{eq:simplicit_elastic_loss}
\end{gathered}
\end{equation}
where $\Psi$ is the strain energy density function depending on the constitutive model (e.g., Neo-Hookean).
The second term, $\mathcal{L}_{\text{ortho}}$ enforces that the skinning weights form an orthonormal basis:
\begin{gather}
\mathcal{L}_{\text{ortho}} = \sum_{i,j=1}^m \int_\Omega (\mathbf W_\theta^i(\mathbf X) \mathbf W_\theta^j(\mathbf X) - \delta_{ij})^2 d\mathbf X.
\end{gather}
This orthogonality constraint not only prevents the trivial solution $\mathbf W \equiv 0$, but also ensures a nice numerical condition for the mass matrix which plays a crucial role in solving Eq.~(\ref{eq:simplicit_sim}) by Newton's method.

\begin{figure}[t]
    \centering
    \includegraphics[width=\columnwidth]{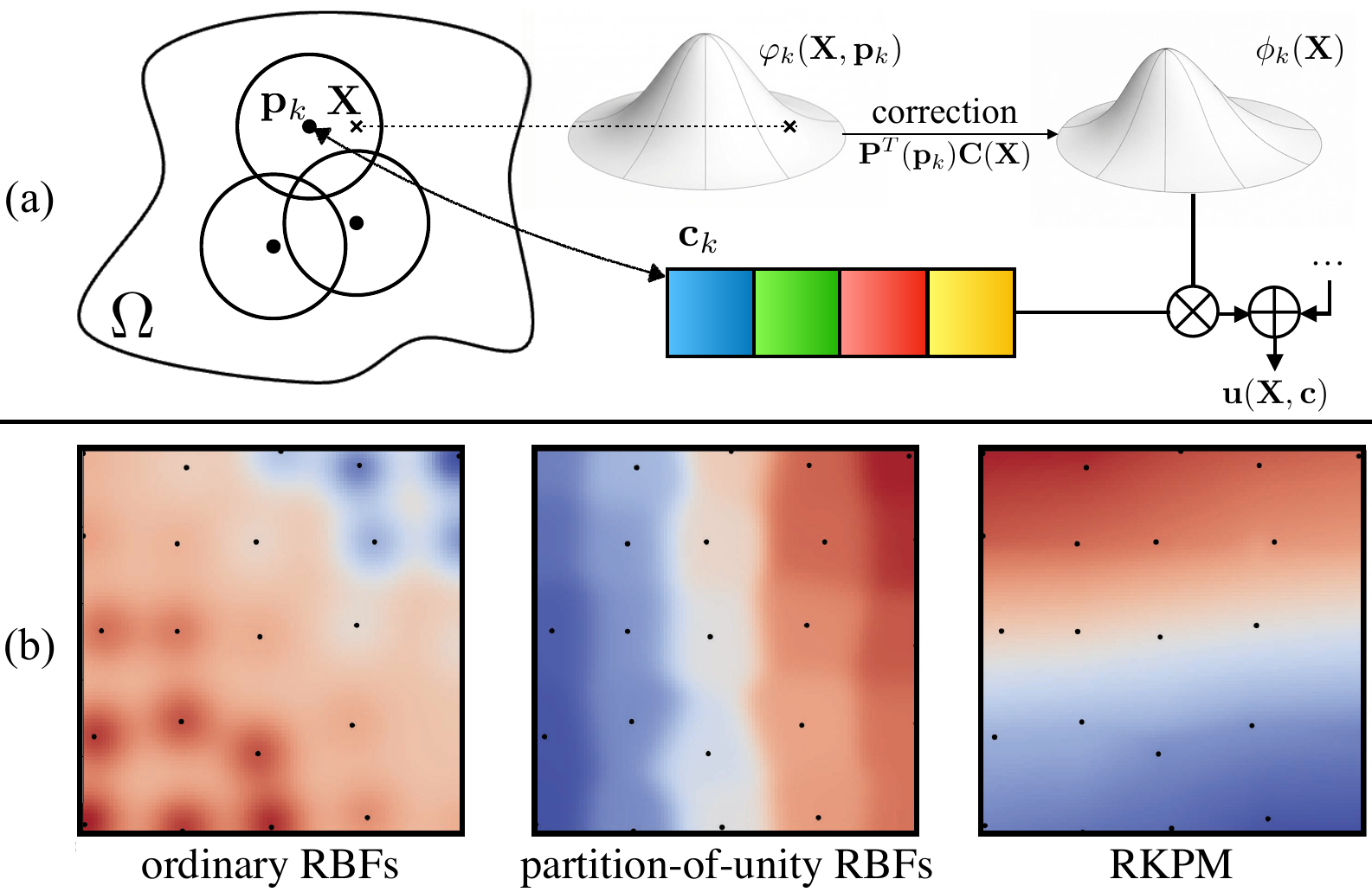}
    \caption{
        (a) Illustration of RKPM. RKPM smoothly interpolates nodal values $\mathbf c_k$ from node centers $\mathbf p_k$ to any query location $\mathbf X$ using the reproducing kernels $\phi_k$ corrected on top of the raw RBF $\varphi_k$ to satisfy the reproducing condition.
        (b) Choosing a suitable kernel basis is essential for high-quality reduced particle simulation.
        We visualize the first nonzero Laplacian eigenmode on three different particle bases.
        Even with nicely-sampled centers, RBFs (\emph{left}) and partition-of-unity RBFs (\emph{middle}) have irregular nonsmooth modes, while RKPMs (\emph{right}, ours) closely approximate the expected linear field.
    }
    \label{fig:rkpm_figure}
\end{figure}

\subsection{Efficient Skinning Eigenmode from RKPM}
\label{sec:method_rkpm}

As an alternative to neural skinning weights, we propose to discretize $\mathbf W$ using the Reproducing Kernel Particle Method (RKPM)~\cite{liu1995reproducing}, which has several key benefits. We first recall the formulation of RKPM (Fig.~\ref{fig:rkpm_figure}), and then show how this representation allows us to efficiently build a high-quality reduced-order basis for elastic deformation.

\paragraph{RKPM} represents any vector-valued function $\mathbf u: \Omega \subset \mathbb R^3 \rightarrow \mathbb R^d$ by a sum of node values $\mathbf c = \{\mathbf c_k \in \mathbb R^d\}_{k=1}^K$ weighted by reproducing kernels $\{\phi_k\}_{k=1}^K$ centered at positions $\{\mathbf p_k \in \Omega\}_{k=1}^K$,
\begin{gather}
    \mathbf u(\mathbf X; \mathbf c) = \sum_{k=1}^K \phi_k(\mathbf X) \mathbf c_k.
    \label{eq:method_rkpm}
\end{gather}
These kernels are conceptually similar to standard Radial Basis Functions (RBF), e.g, Gaussian RBF $\varphi_k(\mathbf X) = \exp(-\Vert \mathbf X - \mathbf p_k\Vert^2 / r^2)$, but are augmented with correction terms to satisfy certain numerical conditions:
\begin{gather}
    \phi_k(\mathbf X) = \varphi_k(\mathbf X) \mathbf P^T(\mathbf p_k) \mathbf C(\mathbf X) ,
    \label{eq:method_correction}
\end{gather}
where $\mathbf P(\mathbf X) = [1, x, y, z]^T$ includes monomials of $\mathbf X$ up to degree $D = 1$ in our case, and $\mathbf C : \Omega \mapsto \mathbb R^{\text{dim} \mathbf P}$ is an introduced correction function to be solved for. The \textit{reproducing condition} requires that these kernels reproduce polynomial functions up to degree $D$ in Eq.~(\ref{eq:method_rkpm}):
\begin{gather}
    \sum_{k=1}^K \phi_k(\mathbf X) \mathbf P(\mathbf p_k) = \mathbf P(\mathbf X)
    \label{eq:method_reproducing}
\end{gather}
Substituting Eq.~(\ref{eq:method_correction}) into Eq.~(\ref{eq:method_reproducing}) yields a linear equation that allows us to solve for $\mathbf C(\mathbf X)$ for any query location $\mathbf X$ as
\begin{gather}
    \mathbf M(\mathbf X)\mathbf C(\mathbf X) = \mathbf P(\mathbf X), \\
    \text{where} \quad \mathbf M(\mathbf X) = \sum_{k=1}^K \varphi_k(\mathbf X) \mathbf P(\mathbf p_k) \mathbf P^T(\mathbf p_k).
\end{gather}
Thus the final formulation of RKPM is
\begin{equation}
\begin{gathered}
    \mathbf u(\mathbf X; \mathbf c) = \sum_{k=1}^K \overbrace{\mathbf P^T(\mathbf p_k)\mathbf C(\mathbf X) \varphi_k(\mathbf x)}^{\phi_k(\mathbf X)} \mathbf c_k \\
    \text{where} \quad \mathbf C(\mathbf X) = \mathbf M^{-1}(\mathbf X)\mathbf P(\mathbf X).
    \label{eq:method_rkpm_final}
\end{gathered}
\end{equation}


\paragraph{Skinning Eigenmode with RKPM.} Directly applying the RKPM formulation in Eq.~(\ref{eq:method_rkpm_final}) to parameterize the deformation map $\Phi(\mathbf X) = \mathbf u(\mathbf X, \mathbf d) + \mathbf X$ allows us to evaluate the elastic potential as
\begin{gather}
    E_\text{pot}^\text{full}(\mathbf d) = \int_\Omega \Psi(\mathbf u(\mathbf X, \mathbf d) + \mathbf X) d\mathbf X,
    \label{eq:method_rkpm_elastic_potential}
\end{gather}
where the DoFs are 3D nodal displacements $\mathbf d = \{\mathbf d_k \in \mathbb R^3\}_{k=1}^K$. Simulating this full-order formulation directly for a large number of RKPM kernels would be costly. In line with existing reduced-order methods, we adopt the skinning deformation in Eq.~(\ref{eq:method_skinning}) and propose to discretize the skinning weight fields $\{\mathbf W^j\}_{j=1}^m$, rather than the displacement $\mathbf u$, with RKPM. The problem boils down to determining adequate nodal values $\mathbf c = [\mathbf c_1, \dots, \mathbf c_K]^T \in \mathbb R^{K \times m}$ for each skinning function $\mathbf W^j(\mathbf X) = \sum_{k=1}^K \phi_k(\mathbf X) \mathbf c_k^j$, where $\mathbf c_k^j$ is the $j$-th element of the vector $\mathbf c_k \in \mathbb R^m$.

Adapting RKPM allows us to extend the Skinning Eigenmode approach~\cite{benchekroun2023fast} to the mesh-free domain. The basic idea is to approximate the elastic potential $E_\text{pot}^\text{full}(\cdot)$ by its Hessian matrix $\mathbf H$ around the rest positions
\begin{gather}
    E_\text{pot}^\text{full}(\mathbf d) \approx \frac12 \mathbf d^T \mathbf H \mathbf d,
\end{gather}
where $\mathbf d \in \mathbb R^{3K}$ is nodal displacements flattened to a vector. Then a set of most expressive skinning weights $\mathbf W$ should be selected so that different columns of $\mathbf c$ minimize the sum of this quadratic form while keeping different channels of $\mathbf W$ orthogonal to each other. Following Benchekroun et al.~\cite{benchekroun2023fast}, we use the simplified weight-space Hessian that prioritizes translation for skinning eigenmode: $\mathbf H_w = \mathbf H_{xx} + \mathbf H_{yy} + \mathbf H_{zz}$, which are blocks of $\mathbf H$ with respect to $x,y,z$ coordinates, respectively. For orthogonality, we have
\begin{equation}
\begin{gathered}
    \delta_{ij} = \langle \mathbf W^i, \mathbf W^j \rangle = \int_\Omega \mathbf W^i(\mathbf X) \mathbf W^j(\mathbf X) d\mathbf X \\
    = \sum_{k,l=1}^K \mathbf c_k^i \mathbf c_l^j \overbrace{\int_\Omega \phi_l(\mathbf X) \phi_m(\mathbf X) d\mathbf X}^{\mathcal M_{lm}} = \sum_{k,l=1}^K \mathbf c_k^i \mathcal M_{lm} \mathbf c_l^j 
    \label{eq:method_orthogonality}
\end{gathered}
\end{equation}
where $\mathcal M$ is the mass matrix of RKPM. Eq.~(\ref{eq:method_orthogonality}) can be written as $\mathbf c^T \mathcal M \mathbf c = \mathbf I$ in matrix form. Putting these together, we have the following optimization problem:
\begin{gather}
\argmin_{\mathbf c \in \mathbb R^{K \times m}} \mathbf \tr(\mathbf c^T \mathbf H_w \mathbf c), \quad \text{subject to} \quad \mathbf c^T \mathcal M \mathbf c = \mathbf I
\label{eq:method_rkpm_optimization}
\end{gather}
This problem can be efficiently solved as a generalized eigenvalue problem $\mathbf H_w \mathbf v = \lambda \mathcal M \mathbf v$, and we can use the first $m$ generalized eigenvectors $[\mathbf v_1, \dots, \mathbf v_m]$ as $\mathbf c$.

\paragraph{Simple expression for Hessian matrix.} We derive a simple expression for the weight-space Hessian matrix $\mathbf H_w$ for the commonly used Neo-Hookean elastic energy that allows easy analytical evaluation. We use the following version of Neo-Hookean elastic energy (Eq. (11) in \cite{smith2018stable})
\begin{gather}
    \Psi(\mathbf F) = \frac12 \left[\bar \lambda (\det \mathbf F - \gamma)^2 + \bar \mu \tr(\mathbf F^T \mathbf F) - E_0\right],
    \label{eq:method_neohookean}
\end{gather}
where $\mathbf F = \nabla \Phi$ is the deformation gradient, $\bar \lambda = \lambda + \mu$ and $\bar \mu = \mu$ are reparametrization of Lamé coefficients $\lambda$ and $\mu$ in linear elasticity (Sec. 3.4 in \cite{smith2018stable}), $\gamma = 1 + {\bar \mu}/{\bar \lambda}$, and $E_0$ is a constant term so that $\Psi(\mathbf I) = 0$.

\begin{prop}
\label{prop:hessian}
For the Neo-Hookean elastic energy above, the $(i,j)$-th element of the weight-space Hessian matrix $\mathbf H_w$ with RKPM discretization in Eq.~(\ref{eq:method_rkpm}) simplifies to
\begin{gather*}
    (\mathbf H_w)_{ij} = \int_\Omega (\lambda(\mathbf X) + 4\mu(\mathbf X)) \nabla \phi_i(\mathbf X)^T \nabla \phi_j(\mathbf X) d\mathbf X.
\end{gather*}
\end{prop}
The proof is given in the supplementary material, and similar expressions can be derived for other commonly used elastic energy functions. Moreover, for homogeneous materials where $\lambda$ and $\mu$ are constant across the domain, the Hessian matrix simplifies to
\begin{equation}
\begin{gathered}
    (\mathbf H_w)_{ij} = (\lambda + 4\mu) \overbrace{\int_\Omega \nabla \phi_i(\mathbf X)^T \nabla \phi_j(\mathbf X) d\mathbf X}^{\mathbf L_{ij}},
\end{gathered}
\end{equation}
where $\mathbf L$ is the weak-form Laplace matrix of RKPM. In this case, the elastic Hessian $\mathbf H_w$ shares the same eigenmodes as the Laplace matrix $\mathbf L$, which is known to be the minimizer of the Dirichlet energy for any input scalar fields. In the heterogeneous case, our proposed RKPM-based skinning eigenmode can be regarded as a material-aware Laplace eigenmode, and connects back to this important concept in classical modal analysis.

\section{Evaluation}
\label{sec:evaluation}

\begin{figure*}[t]
    \centering
    \includegraphics[width=\textwidth]{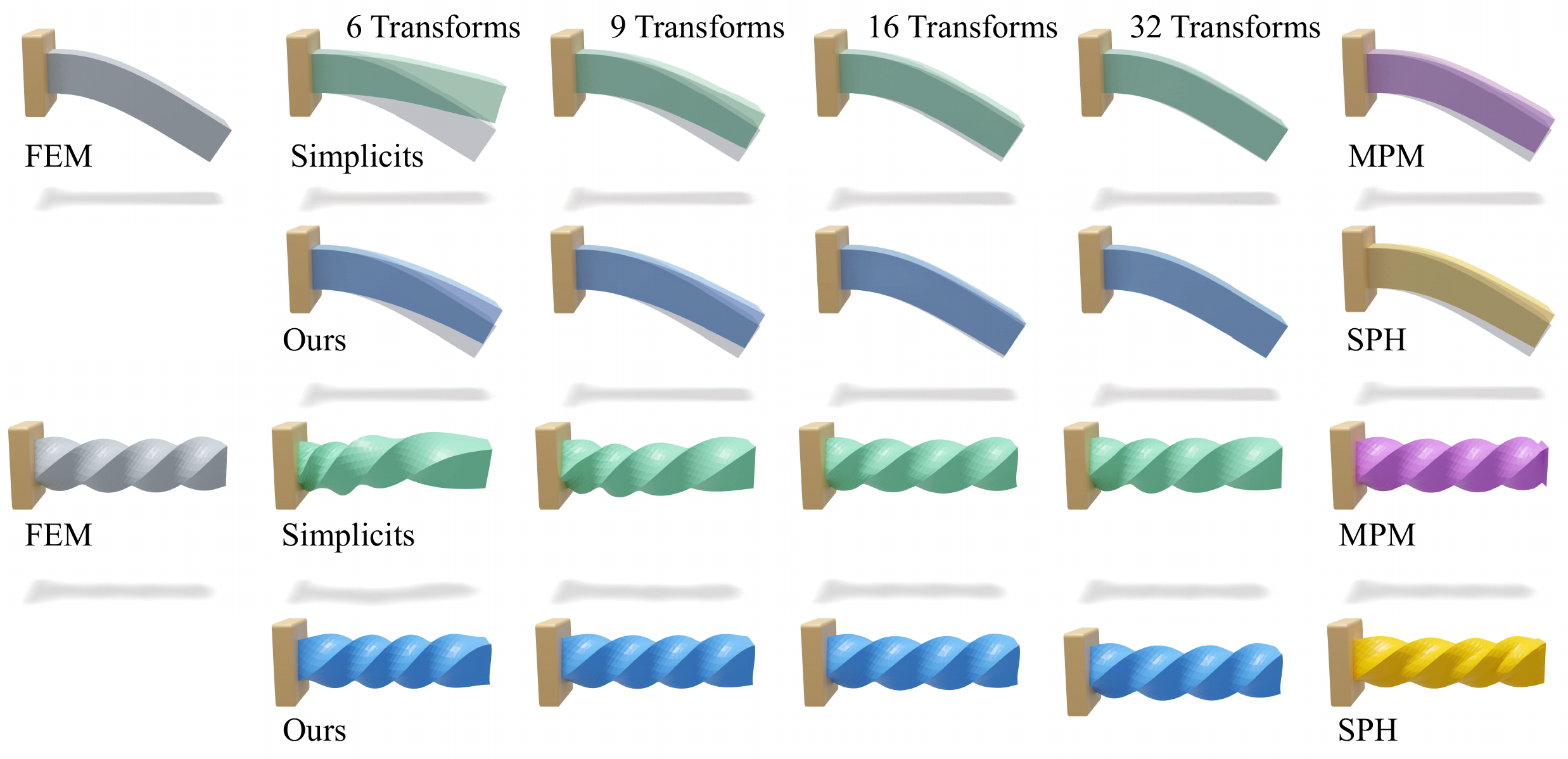}
    \caption{
    Visual comparison on standard beam test. For the case of bending cantilever beam, FEM solution is overlaid semi-transparently on top of the simulated result from all competing methods to aid visual comparison.}
    \label{fig:result_beam}
\end{figure*}


\begin{table}[t]
  \centering
  \begin{tabular}{l l c c c c}
  \toprule
Test & $m$ & Simplicits & Ours & MPM & SPH \\
  \midrule
\multirow{4}{*}{Bend} & 6 & 1.20e-02 & 7.80e-03 & \multirow{4}{*}{1.42e-03} & \multirow{4}{*}{6.57e-04} \\ 
                      & 9 & 6.94e-03 & 4.90e-03 & & \\ 
                      & 16 & 1.53e-03 & 4.10e-04 & & \\ 
                      & 32 & 1.17e-04 & \textbf{2.93e-06} & & \\ 
  \midrule
\multirow{4}{*}{Twist} & 6 & 2.54e-03 & 1.56e-04 & \multirow{4}{*}{2.34e-05} & \multirow{4}{*}{1.33e-04} \\ 
& 9 & 3.42e-04 & 2.95e-05 & & \\ 
& 16 & 1.30e-04 & \textbf{3.46e-06} & & \\ 
& 32 & 4.21e-05 & 6.64e-06 & & \\ 
  \bottomrule
  \end{tabular}
  \caption{Quantitive evaluation on the standard beam deformation test. We report the normalized Mean Squared Error (MSE) of simulated point locations on two types of boundary conditions. The results are reported for different numbers $m$ of affine transformations for Simplicits and our method. We also show the results of MPM and SPH for comparison.}
  \label{table:eval_beam}
\end{table}


\paragraph{Standard beam test.} We start with evaluation results on a cuboid beam, a standard test in the simulation of elastodynamics. We follow the same experiment setup as Simplicits~\cite{modi2024simplicits}, using a beam of shape $5\mathrm{m}\times1\mathrm{m}\times1\mathrm{m}$, Young's Modulus $5 \times 10^6 \mathrm{Pa}$, Poisson ratio $0.45$, and density $1 \times 10^3 \mathrm{kg/m^3}$. We apply two types of boundary conditions for the beam: (1) fixing the leftmost $0.5\mathrm{m}$ of the beam and letting the right side bend freely (named `Bend'); (2) fixing the leftmost end of the beam and twisting the right end of the beam by up to $720^{\circ}$ (named `Twist'). 

We compare our method with three other types of mesh-free methods, Simplicits~\cite{modi2024simplicits}, MPM and SPH, against simulation results with Finite Element Methods (FEM) on tetrahedral meshes, which is regarded as the gold standard for elastodynamic simulation. We use the Mean Squared Error (MSE) of simulated point locations across the whole shape and all frames, normalized by the bounding box size of the reference shape. Simplicits and ours are reduced-order methods, so we also report results with different numbers $m$ of skinning functions or affine transformations. The experiment results are reported in Table \ref{table:eval_beam} and shown in Fig.~\ref{fig:result_beam}.

With both boundary conditions, Simplicits and our method show steady improvement in accuracy as the number of affine transformations increases, meaning an increase in the number of DoFs allowed in the simulation. Our method consistently outperforms Simplicits with the same DoFs. When the number of transformations reaches a certain threshold, our simulation results can match or surpass the accuracy of MPM and SPH, two full-order simulation methods with different formulations.

\begin{table*}[t]
  \centering
  \begin{tabular}{c c c c c c c c}
  \toprule
\multirow{2}{*}{Method} & \multicolumn{2}{c}{Fix Side} & \multicolumn{2}{c}{Pull Farthest} & \multicolumn{2}{c}{Pull Boundary} & \multirow{2}{*}{Training time (s)} \\
& MSE & Max & MSE & Max & MSE & Max \\
    \midrule 
Simplicit & 8.97e-03 & 2.64e-02 & 5.58e-02 & 1.66e-01 & 3.37e-02 & 6.30e-02 & 121.44 $\pm$ 10.15 \\
Ours & \textbf{6.87e-03} & \textbf{2.14e-02} & \textbf{3.75e-02} & \textbf{1.19e-01} & \textbf{3.11e-02} & \textbf{5.96e-02} & \textbf{3.19} $\pm$ 2.48\\
Improvement & 34.2\% & 18.2\% & 29.8\% & 27.3\% & 37.5\% & 38.9\% & 97.4\% \\
\midrule
Simplicit & 2.16e-09 & 2.55e-09 & 9.38e-04 & 2.77e-03 & 8.83e-04 & 2.60e-03 & 117.45 $\pm$ 1.13 \\
Ours & \textbf{1.01e-09} & \textbf{1.29e-09} & \textbf{4.75e-04} & \textbf{1.40e-03} & \textbf{4.16e-04} & \textbf{1.26e-03} & \textbf{3.49} $\pm$ 2.39 \\
Improvement & 18.9\% & 18.3\% & 38.1\% & 40.1\% & 45.4\% & 44.4\% & 97.0\% \\
  \bottomrule
  \end{tabular}
  \caption{Quantitive evaluation on the Thingi10K and Simready Datasets. We report the normalized Mean Squared Error (MSE) and maximum error across all the examples for each boundary condition. We also show the training time of our method compared to Simplicits, and the improvement of our method in percentage, for a total of $m=32$ skinning functions.}
  \label{table:eval_thingi10k_simready}
\end{table*}

\begin{figure*}[t]
    \centering
    \includegraphics[width=0.88\textwidth]{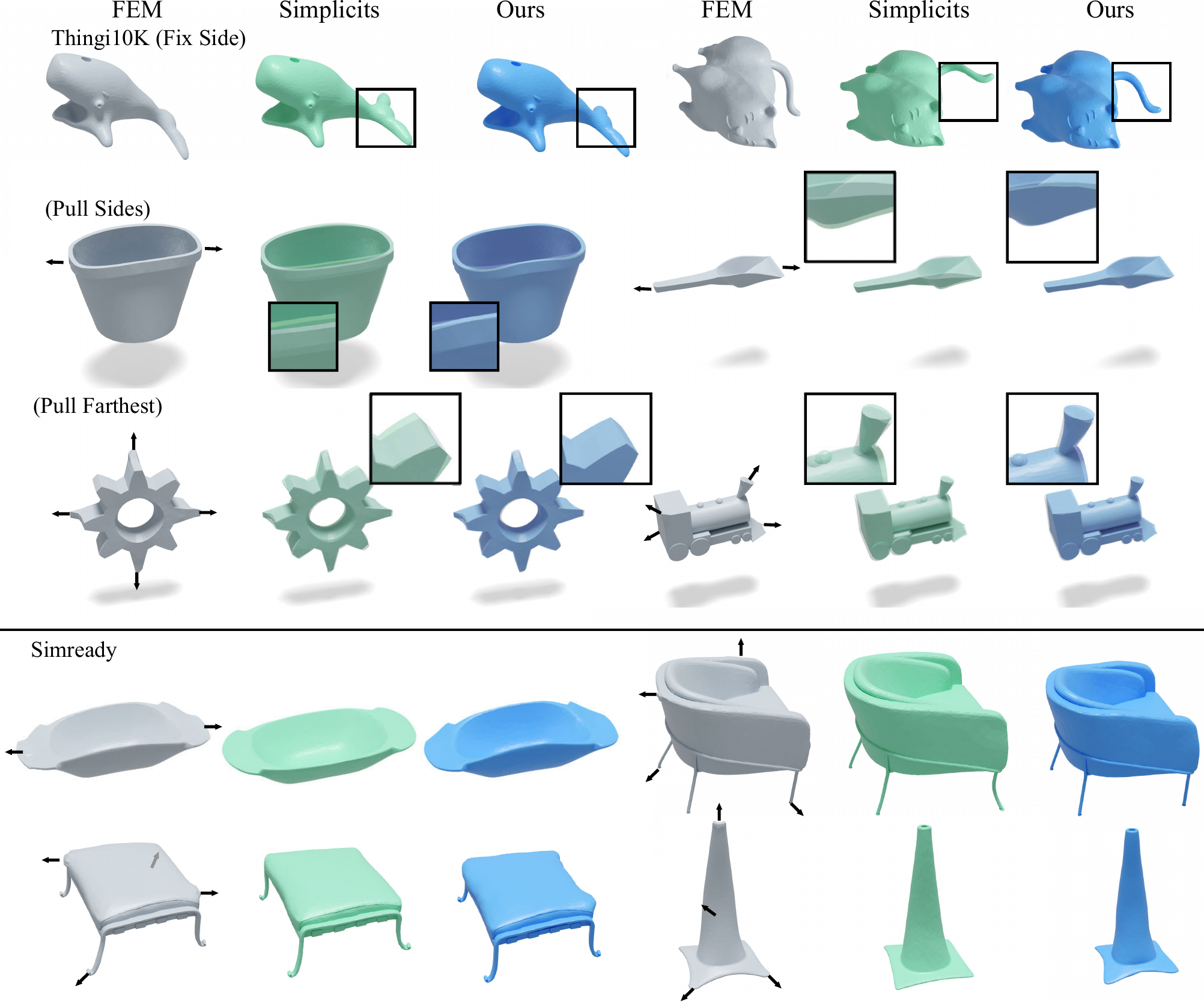}
    \caption{
    Comparison on Thingi10k and Simready dataset. In ``Thingi10K (Fix Side)'', the leftmost sides of the objects are fixed. In ``Pull Sides'' and ``Pull Farthest'', black arrows around the FEM results in grey visualize the locations and directions of applied moving boundary conditions, and we overlay FEM results on top of the results from Simplicits in green and our method in blue, along with zoom-in views to highlight the discrepancy from the converged FEM solution.
    See Table~\ref{table:eval_thingi10k_simready} and the attached video for the increased fidelity of our simulations.
    }
    \label{fig:result_thingi10k}
\end{figure*}


\paragraph{Thingi10K and Simready datasets.} To evaluate our method on more diverse input examples, we take 20 shapes from the Thingi10K dataset~\cite{zhou2016thingi10k} and 19 shapes from the Simready Dataset~\cite{simready}. The shapes we select need to satisfy several geometric properties, such as being manifold, oriented, without self-intersection, and enclosing clear volumes in order to obtain tetrahedral meshes for FEM simulation ground truth. We use TetWild~\cite{hu2018tetrahedral} for the tetrahedralization of those filtered shapes from both datasets. In this experiment, we focus on comparing our method with Simplicits~\cite{modi2024simplicits}, since it is the only method with the same problem setting as ours.

The Thingi10K dataset is an online collection of 3D models covering a wide range of categories including both imaginary and real-world objects. The shapes come in different scales, so we normalize their bounding box sizes to be $1$. Depending on the semantic meaning of the object, we manually assign Young's Modulus of around $10^5\mathrm{Pa}$ to organic shapes and $10^8\mathrm{Pa}$ to other categories, while keeping Poisson ratio and density consistent across the dataset.

The Simready dataset consists of meshes of real-world objects created by artists, where the shapes are provided in metric scales. We utilize VoMP~\cite{dagli2026vomp} to predict the volumetric physical parameters including Young's Modulus, Poisson ratio, and density, and use these properties for both the FEM ground truth simulation and the methods in comparison.

We report the simulation results in three types of boundary conditions: (1) fixing one side of the objects; (2) pulling the objects in different directions on 4 farthest points sampled from the surface; (3) pulling the objects on two sides along its longest axis. The results are reported in Table \ref{table:eval_thingi10k_simready} and Fig.~\ref{fig:result_thingi10k}. To evaluate simulation accuracy, we report the normalized Mean Squared Error (MSE) and the maximum displacement error in each simulation sequence averaged over all the examples in the dataset. We also compare the average training time of our method, including the computation of Hessian matrix and mass matrix followed by eigen-decomposition in Eq.~(\ref{eq:method_rkpm_optimization}), with Simplicits. Our method achieves consistently better simulation accuracy than Simplicits, and achieves around 40x faster training speed than Simplicits, thanks to our eigenanalysis formulation.

\begin{table}[t]
  \centering
  \begin{tabular}{l l c c c}
  \toprule
Loss type & Sampling & Drop & Twist & Time (s) \\
  \midrule
Simplicits & (Random) & 1.53e-3 & 1.30e-3 & 114.28 \\
  \midrule
Random $\mathbf z$ & Random & 1.58e-2 & 2.50e-2 & 160.12 \\
Random $\mathbf z$ & Grid & 1.24e-2 & 7.29e-3 & 412.38 \\
  \midrule
Hessian & Random & 4.86e-3 & 5.07e-4 & 103.66 \\
Hessian & Grid & 4.45e-4 & 3.49e-5 & 145.96 \\
  \midrule
Ours & (Grid) & \textbf{4.10e-4} & \textbf{3.46e-5} & \textbf{3.93} \\
  \bottomrule
  \end{tabular}
  \caption{Ablation results on different training strategy. We compare variants of our method trained using different loss functions and integration point sampling methods. We also highlight the efficiency of our eigenanalysis formulation over gradient-based optimization in terms of training time for $m=32$.}
  \label{table:eval_ablation_training}
\end{table}

\paragraph{Ablation study on training.} We conduct ablation studies on differences in training strategy between our method and Simplicits to justify our design choices. We use the standard beam test in Table \ref{table:eval_beam}, where all competing methods are allowed $m = 16$ affine transformations, and use the RKPM discretization except the original Simplicits method. All models, except ours, are trained using iterative optimization for the same number of iterations. The results are reported in Table \ref{table:eval_ablation_training}.

Besides using neural field or RKPM, two other key differences between Simplicts and our method are the loss variants and point sampling method for training. We first compare the expected elastic energy over randomly sampled transformations used in Simplicits in Eq.~(\ref{eq:simplicit_elastic_loss}) with the Hessian approximation in Eq.~(\ref{eq:method_rkpm_optimization}) in our method. The results show that the Hessian loss achieves a better simulation accuracy for RKPM parameterization. Second, for the Monte Carlo integration of elastic energy in Eq.~(\ref{eq:simplicit_elastic_loss}) and (\ref{eq:method_rkpm_elastic_potential}), we compare the strategy of sampling random points in different iterations, as in Simplicits, with using the same points sampled from a uniform grid, as in our method. The results show that our uniform grid sampling strategy performs better given sufficient grid resolution.

Lastly, our method has almost the same formulation as the ``Hessian - Grid'' baseline in the table, only except that the baseline uses a gradient-based iterative optimization of loss terms like Eq.~(\ref{eq:simplicit_loss_sum}), whereas our method solves a generalized eigenvalue problem using highly efficient linear algebra routines. As a result, these two variants are on par in accuracy, while our method uses a significantly lower training time. In addition, the optimization in Eq.~(\ref{eq:simplicit_loss_sum}) imposes orthogonality as a soft penalty constraint, whereas the output skinning weights $\mathbf W$ from eigen-decomposition satisfy exact orthogonality (up to numerical precision). This is beneficial to the numerical condition of the system Hessian during simulation (Eq.~(\ref{eq:simplicit_sim})).

\paragraph{Ablation study on test-time sampling.} Simplicits randomly samples integration points inside the object domain in different training iterations, while our method computes the elastic energy on a limited set of sample points due to the eigen-decomposition. As a result, one may wonder whether our simulation result is more sensitive to different integration samples at test time. In Table \ref{table:eval_ablation_sampling}, we show a comparison of our method and Simplicits using 5k points from a uniform grid vs. randomly sampled points. It turns out that although both Simplicits and our method both show variation in accuracy due to sampling differences, our method still achieves lower error in both cases.

\begin{table}[t]
  \centering
  \begin{tabular}{l l c c}
  \toprule
Test & Sampling & Simplicits & Ours \\
  \midrule
\multirow{2}{*}{Drop} & Grid & 1.53e-3 & \textbf{4.10e-4} \\ 
                       & Random  & 3.69e-3 & 9.42e-4 \\
  \midrule
\multirow{2}{*}{Twist} & Grid & 1.30e-4 & \textbf{3.46e-6} \\ 
                       & Random & 3.88e-4 & 1.14e-5 \\
  \bottomrule
  \end{tabular}
  \caption{Ablation studies on sampling method of integration points in simulation stage. We test with 5k integration points sampled from a uniform grid or random uniform distribution.}
  \label{table:eval_ablation_sampling}
\end{table}


\paragraph{Qualitative results.}
We demonstrate qualitative simulation results of our method in several different scenarios Fig.~\ref{fig:teaser}. We first show a set of 3DGS objects dropped on a table individually, and then two large-scale simulation scenes, where 13 different splats being dropped in a container and 18 gaussian splat dog toys falling through a plinko machine, respectively. We also show an application of simulating a robot arm interacting with several 3DGS objects. Our results are best viewed in videos, and please refer to our supplementary materials for more results and detail.

\section{Limitation and Future Work}
Although the explicit RKPM discretization offers clear advantages over implicit neural bases, it also has several limitations. Some are shared with all reduced representations, while others are specific to RKPM. As a reduced-order model, high-frequency details such as wrinkles are difficult to capture because the global basis represents smooth, low-frequency deformation. Large nonlinear effects, including sharp contact, are likewise challenging since the basis is smooth and derived from a linearization around the rest state. As with most reduced models, we do not model topology changes such as fracture by default.
Finally, RKPM depends on kernel radius, sampling density, and particle distribution, which requires careful implementation to ensure basis quality. These limitations suggest several promising directions for future work.


\section*{Acknowledgments}
We thank Sangeetha Grama Srinivasan for helping us with comparisons to SPH. We thank Miles Macklin, Eric Shi, Lukasz Wawrzyniak and others from the Warp team for assistance with the Warp platform. We thank Masha Shugrina, Clement Fuji Tsang, Or Perel and others for helping us with Kaolin and rendering 3DGS with 3DGRUT. We also thank the anonymous reviewers for their helpful feedback.

{
    \small
    \bibliographystyle{ieeenat_fullname}
    \bibliography{main}
}

\clearpage
\setcounter{page}{1}
\maketitlesupplementary

\section{Further Discussion on MPM and SPH}

\label{sec:supp_discussion_mpm_sph}

As mentioned in Sec.~\ref{sec:related}, many particle-based physics simulation methods have been proposed.
MPM and SPH are particularly attractive, as they can handle a wide variety of material models, including plasticity effects, topology and phase changes, and, as is our focus here, elastodynamics.

\begin{figure}[t]
    \centering
    \includegraphics[width=\columnwidth]{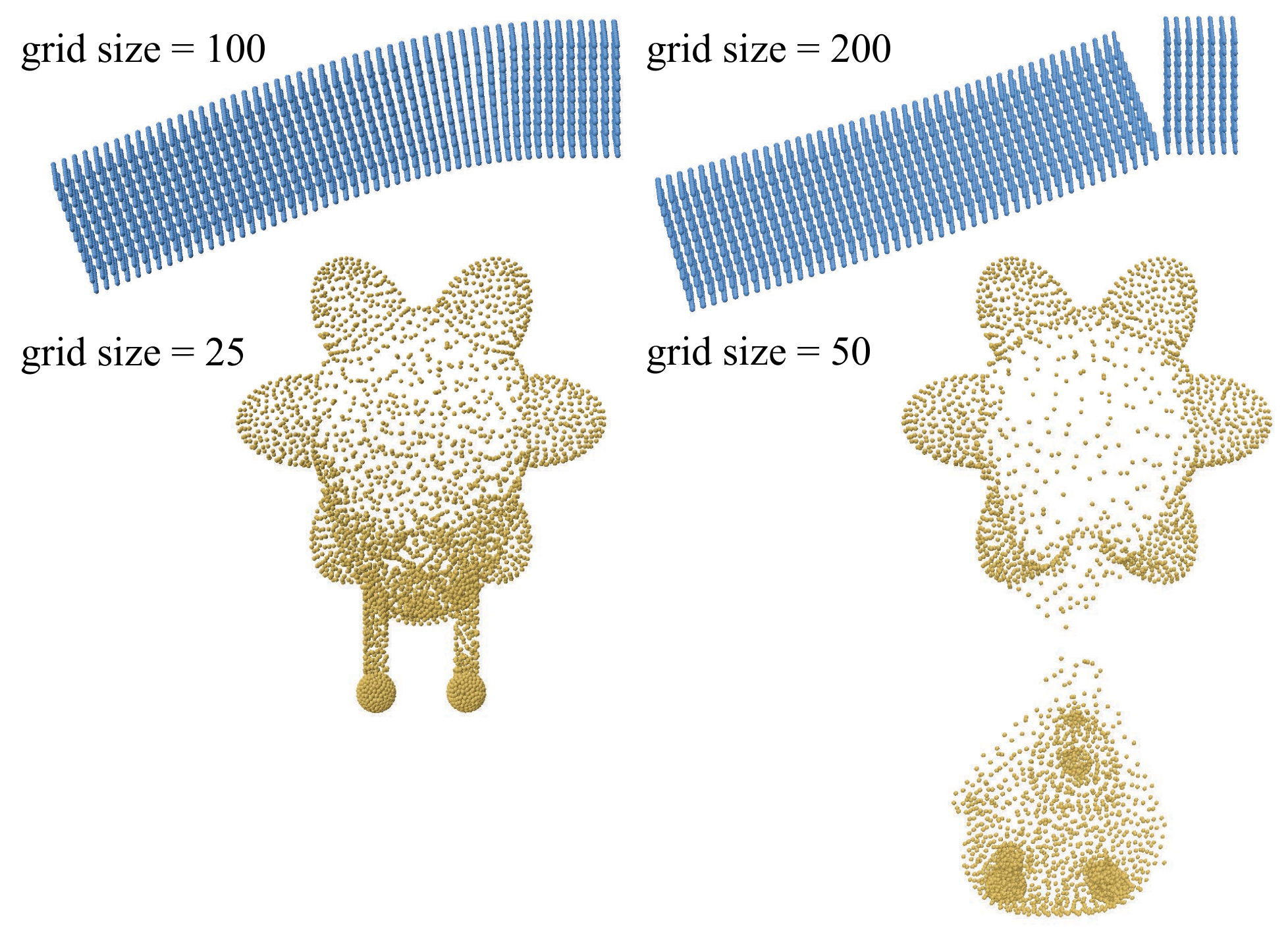}
    \caption{The Material Point Method (MPM) is versatile, but presents challenges for deformable body simulation due to the occurrence of unintended numerical fracture depending on the grid resolution (shown in top-left corner).}
    \label{fig:mpm_fracture}
\end{figure}

However, this versatility regarding topology changes is also what makes MPM and SPH sensitive to spatial discretization, as the interaction stencils change over time. For MPM, two particles will interact if and only if they share at least one common grid node; for SPH, only if the support of their kernels overlap. Inevitably, as the material is increasingly stretched, particles will eventually get further apart than this critical distance, and numerical fracture will happen, as illustrated in Fig.~\ref{fig:mpm_fracture}.
For SPH, particles becoming locally co-dimensional can also lead to numerical conditioning issues, with the velocity gradient becoming singular and requiring special care~\citep{westhofen23sph}. For MPM elastic bodies, deformation gradient estimation can be made more robust by leveraging a rest pose mesh~\citep{Jiang2016mpm}, or even by rasterizing forces from a Lagrangian model~\citep{fei2018wetcloth}, when such a representation is available. However, integration accuracy will still suffer when the number of particles per cell is not sufficient, while a too coarse grid will exhibit locking; this makes picking the grid resolution difficult for uneven particle distributions.

The same sampling criteria apply to our RKPM kernel centers; however, our method only needs to worry about the rest pose, for which it is easier to control the sample distribution and kernel width, while MPM and SPH need to have the particles remain well distributed at each timestep. 
Resampling particles over time can avoid those issues~\citep{yue2015continuum}; however, this is not really practical when simulating a predefined number of Gaussian splats, for instance.

Moreover, while so-called implicit variants of MPM and SPH have been proposed, most still treat advection as an explicit step and are therefore subject to the Courant–Friedrichs–Lewy (CFL) condition. In contrast, our total-Lagrangian approach, with shape functions remaining fixed over time and implicit time stepping, does not have a constraint on the size of time step. 

\section{Implementation Details}

\subsection{Our Method}

\paragraph{RKPM construction.} Given an input object, our method first constructs a set of RKPM kernels around the object shape. We start by dense sampling integration points around the object. Unless otherwise stated, we sample points on a uniform grid inside the object bounding box, and then reject points outside the object for shapes with well-defined an inside/outside test functions, e.g., a watertight mesh. For shapes where the inside/outside test cannot be easily applied, e.g. 3DGS, we directly use the given points as integration points (after downsampling if necessary to avoid out-of-memory error). With the integration points determined, we then apply Farthest Point Sampling (FPS) to select around 1k points as RKPM kernel centers to ensure that kernels are equidistantly distributed. We set each Gaussian kernel radius $r$ to be the minimal distance to reach two other centers, so that the space around the kernels is well-covered by RKPM.

\paragraph{Eigenanalysis.} After RKPM kernels are constructed, we assemble the Hessian matrix $\mathbf H_w$ according to Prop.~\ref{prop:hessian} and $\mathcal M$ according to Eq.~(\ref{eq:method_orthogonality}). We perform generalized eigen-decomposition using \texttt{torch.lobpcg} on CUDA in double precision. We take the eigenvectors associated with $m$ smallest eigenvalues as the nodal values for our skinning weight estimation. Notice that the solutions of the eigen-decomposition always includes a constant mode associated with zero eigenvalue. Simplicits does not find this constant mode in its optimization process, and thus manually append a constant mode that corresponds to a global affine transformation. To ensure a fair comparison with Simplicits, when reporting the number $m$ of skinning weights, we do not include the constant mode for both methods throughout the paper.

\paragraph{Simulation.} Once the skinning weights are determined, we run simulation of deformable objects in the same approach as Simplicits~\cite{modi2024simplicits}. Our implementation is built on top of the open-source Kaolin library based on the Warp language and PyTorch, and we further boost the runtime performance with more efficient kernel launching via CUDA graph captured in Warp and PyTorch while maintaining the same simulation results. In the paper, we test both our method and Simplicits with the same enhanced implementation for fairness. We use Newton's method with line search based on Wolfe conditions to solve implicit time-stepping in Eq.~(\ref{eq:simplicit_sim}), allowing up to 20 updates per time step with a convergence tolerance of $10^{-8}$. To solve the linear system in Newton's method, we use the direct solver \texttt{torch.linalg.solve} in PyTorch.

\subsection{Baseline Methods}

\paragraph{Simplicits.} We use the recommended implementation in Kaolin. The neural field for skinning weights is a 6-layer MLP (excluding the input and output layers) of layer width 64, trained for 10k iterations using the Adam optimizer, with a learning rate of $10^{-3}$. The elasticity and orthogonality loss weights are set to $0.1$ and $10^6$, respectively. At run time, we adopt the same simulation implementation as our method for Simplicits.

\paragraph{MPM and SPH.} For MPM, we use the GPU-based warp-mpm\footnote{\url{https://github.com/zeshunzong/warp-mpm/}} implementation. For the standard beam test, we use $\text{dt} = 10^{-4}\mathrm{s}$, whereas $\text{dt} = 10^{-3}\mathrm{s}$ leads to numerical explosion. For SPH, we use the elasticity model in \citet{kugelstadt2021fast} implemented in SPlisHSPlasH\footnote{\url{https://github.com/InteractiveComputerGraphics/SPlisHSPlasH}}, and $\text{dt} = 0.01\mathrm{s}$ for the beam test.

\paragraph{Finite Element Methods.} Finite Element Methods are widely adopted and regarded as the standard approach for simulating elasticity. In this work, we use converged FEM simulation results as the gold standard reference for evaluating various methods. Our full-DoF FEM simulation is implemented based on \texttt{warp.fem}\footnote{\url{https://nvidia.github.io/warp/modules/fem.html}}. The simulation uses the same Neo-Hookean elasticity model in Eq.~(\ref{eq:method_neohookean}) on tetrahedral meshes. The solver adopts the backward Euler method for implicit time stepping, solved by Newton's method. 

 \begin{table}[t]
    \centering
    \begin{tabular}{l c c c c c}
    \toprule
  $m$ & Simplicits & \textbf{Ours} & FEM & MPM & SPH \\
    \midrule
  6 & 5.01 & 3.01 & \multirow{4}{*}{427.2} & \multirow{4}{*}{23.1} & \multirow{4}{*}{37.8} \\
  9 & 3.95 & 3.71 & & & \\
  16 & 5.09 & 5.42 & & & \\
  32 & 10.0 & 10.7 & & & \\
    \bottomrule
    \end{tabular}
    \caption{Comparison of runtime in milliseconds (ms) for a simulation step of $\text{dt} = 0.01\mathrm{s}$ on average. The timing results are reported for the beam-bending experiment.}
    \label{table:eval_runtime}
\end{table}

\subsection{Runtime Comparison}

We report the time used to simulate a period of $0.01\mathrm{s}$ in the beam-bending experiment by various methods in Table~\ref{table:eval_runtime}. Our method and Simplicits adopt the same solver and therefore reach similar runtime performance. FEM is a full-DoF simulation that yields accurate results but takes orders of magnitude longer to run. MPM and SPH also achieve competitive runtime performance, but are still slower than our reduced-order formulation.

\section{Proof of Proposition~\ref{prop:hessian}}

In this section, we provide a proof of Proposition~\ref{prop:hessian}. Consider the deformation map $\Phi(\mathbf X, \mathbf d) = \mathbf u(\mathbf X, \mathbf d) + \mathbf X$, where the displacement $\mathbf u(\mathbf X, \mathbf d)$ is parameterized by RKPM in Eq.~(\ref{eq:method_rkpm}) and the DoFs are the nodal displacements $\mathbf d = \{\mathbf d_k \in \mathbb R^3\}_{k=1}^K = \{(\mathbf d_k^x, \mathbf d_k^y, \mathbf d_k^z)^T\}_{k=1}^K$.
\begin{equation}\begin{gathered}
    \Phi(\mathbf X, \mathbf d) = \begin{bmatrix} \Phi^x(\mathbf X, \mathbf d) \\ \Phi^y(\mathbf X, \mathbf d) \\ \Phi^z(\mathbf X, \mathbf d) \end{bmatrix} \\
    = \sum_{k=1}^K \phi_k(\mathbf X) \mathbf d_k + \mathbf X,
\end{gathered}\end{equation}
The deformation gradient $\mathbf F(\mathbf X, \mathbf d) \in \mathbb R^{3 \times 3}$ is the spatial derivative of $\Phi$ at each point $\mathbf X \in \Omega$, given by
\begin{equation}\begin{gathered}
    \mathbf F(\mathbf X, \mathbf d) = \nabla \Phi(\mathbf X, \mathbf d) = \begin{bmatrix} \nabla \Phi^x(\mathbf X, \mathbf d)^T \\ \nabla \Phi^y(\mathbf X, \mathbf d)^T \\ \nabla \Phi^z(\mathbf X, \mathbf d)^T \end{bmatrix} \\
    = \sum_{k=1}^K \mathbf d_k \nabla \phi_k(\mathbf X)^T  + \mathbf I \\
    = \overbrace{\begin{bmatrix} \mathbf d_1 & \dots & \mathbf d_K \end{bmatrix}}^{\mathbf d^T \in \mathbb R^{3 \times K}} \overbrace{\begin{bmatrix} \nabla \phi_1(\mathbf X)^T \\ \vdots \\ \nabla \phi_K(\mathbf X)^T \end{bmatrix}}^{\nabla \boldsymbol\phi(\mathbf X) \in \mathbb R^{K \times 3}} + \mathbf I.
\end{gathered}\end{equation}
where $\nabla$ denotes the derivative with respect to the point $\mathbf X$. With a slight abuse of notation, we write
\begin{equation}\begin{gathered}
    \boldsymbol\phi(\mathbf X) = \begin{bmatrix} \phi_1(\mathbf X) \\ \dots \\ \phi_K(\mathbf X) \end{bmatrix} \in \mathbb R^K, \\
    \mathbf d = \begin{bmatrix} \mathbf d_1^T \\ \vdots \\ \mathbf d_K^T \end{bmatrix} = \begin{bmatrix} \mathbf d^x, \mathbf d^y, \mathbf d^z \end{bmatrix} \in \mathbb R^{K \times 3},
\end{gathered}\end{equation}
where $\mathbf d^x, \mathbf d^y, \mathbf d^z \in \mathbb R^K$ are the nodal displacements in the $x,y,z$ directions, respectively. Then we have
\begin{equation}\begin{gathered}
    \mathbf F(\mathbf X, \mathbf d) = \begin{bmatrix} \mathbf F_{11} & \mathbf F_{12} & \mathbf F_{13} \\ \mathbf F_{21} & \mathbf F_{22} & \mathbf F_{23} \\ \mathbf F_{31} & \mathbf F_{32} & \mathbf F_{33} \end{bmatrix} \\
    = \begin{bmatrix} (\mathbf d^x)^T \\ (\mathbf d^y)^T \\ (\mathbf d^z)^T \end{bmatrix} \begin{bmatrix} \partial_x \boldsymbol\phi(\mathbf X) & \partial_y \boldsymbol\phi(\mathbf X) & \partial_z \boldsymbol\phi(\mathbf X) \end{bmatrix} + \mathbf I \\
    = \mathbf d^T \nabla \boldsymbol\phi(\mathbf X) + \mathbf I,
    \label{eq:supp_F}
\end{gathered}\end{equation}
where $\partial_x \boldsymbol\phi(\mathbf X) = \begin{bmatrix} \partial \phi_1(\mathbf X) / \partial x & \dots & \partial \phi_K(\mathbf X) / \partial x \end{bmatrix}^T$, and similarly for $\partial_y \boldsymbol\phi(\mathbf X)$ and $\partial_z \boldsymbol\phi(\mathbf X)$.

Applying the strain energy density and integrating over the domain $\Omega$ by the Monte Carlo method, we obtain the total elastic potential energy as
\begin{equation}\begin{gathered}
    E_\text{pot}(\mathbf d) = \int_\Omega \Psi(\mathbf F(\mathbf X, \mathbf d)) d\mathbf X \approx \sum_{i} v_i \Psi(\mathbf F(\mathbf X_i, \mathbf d)),
    \label{eq:supp_elastic_potential_energy}
\end{gathered}\end{equation}
where $\mathbf X_i$ is the $i$-th sample point, and $v_i$ is the weight of the $i$-th sample point. Then its weight-space Hessian matrix
\begin{equation}\begin{gathered}
    \mathbf H_w = \mathbf H_{xx} + \mathbf H_{yy} + \mathbf H_{zz}
\end{gathered}\end{equation}
contains the Hessian of $E_\text{pot}$ with respect to the nodal displacements $\mathbf d^x, \mathbf d^y, \mathbf d^z$ around the rest position $\mathbf d = \mathbf 0$, respectively. Take $\mathbf H_{xx}$ as an example, and denote $\Hess(\cdot, \cdot)$ as the Hessian of the first argument with respect to the second argument.
\begin{equation}\begin{gathered}
    \mathbf H_{xx} = \Hess(E_\text{pot}, \mathbf d^x) = \Hess(\sum_{i} v_i \Psi(\mathbf F_i), \mathbf d^x) \\
    = \sum_{i} v_i \Hess(\Psi(\mathbf F_i), \mathbf d^x),
    \label{eq:supp_hess_dx}
\end{gathered}\end{equation}
where $\mathbf F_i = \mathbf F(\mathbf X_i, \mathbf d)$ is a shorthand for the deformation gradient at the $i$-th sample point. Notice that $\mathbf F$ is an affine transformation of $\mathbf d$, so the chain rule for Hessian matrix gives
\begin{equation}\begin{gathered}
    \Hess(\Psi(\mathbf F_i), \mathbf d^x) = \mathbf J_i^T \Hess(\Psi(\mathbf F_i), \mathbf F_i) \mathbf J_i,
\end{gathered}\end{equation}
where $\mathbf J_i$ is the Jacobian matrix of $\Vector(\mathbf F_i)$ with respect to $\mathbf d^x$ around $\mathbf d = 0$ and $\Hess(\Psi(\mathbf F_i), \mathbf F_i)$ is the Hessian of strain energy density with respect to the flattened $\Vector(\mathbf F_i) \in \mathbb R^9$ around $\mathbf F_i = \mathbf I$. For the Neo-Hookean energy in Eq.~(\ref{eq:method_neohookean}), we have the gradient
\begin{equation}\begin{gathered}
    \grad(\Psi(\mathbf F_i), \mathbf F_i) = \mu \mathbf F_i + (\lambda + \mu) (J - \gamma) J\mathbf F_i^{-T}, \\
    J = \det(\mathbf F_i)
\end{gathered}\end{equation}
and the Hessian
\begin{equation}\begin{gathered}
    \Hess(\Psi(\mathbf F_i), \mathbf F_i) = \mu \mathbf I \\
    + (\lambda + \mu)(2J - \gamma) \Vector(J\mathbf F_i^{-T})^T \Vector(\mathbf F_i^{-T}) \\
    - (\lambda + \mu)(J - \gamma)J(\mathbf F_i^{-1} \otimes \mathbf F_i^{-T}) \mathbf K, \\
    \mathbf K = \begin{bmatrix}
        1 & 0 & 0 & 0 & 0 & 0 & 0 & 0 & 0 \\
        0 & 0 & 0 & 1 & 0 & 0 & 0 & 0 & 0 \\
        0 & 0 & 0 & 0 & 0 & 0 & 1 & 0 & 0 \\
        0 & 1 & 0 & 0 & 0 & 0 & 0 & 0 & 0 \\
        0 & 0 & 0 & 0 & 1 & 0 & 0 & 0 & 0 \\
        0 & 0 & 0 & 0 & 0 & 0 & 0 & 1 & 0 \\
        0 & 0 & 1 & 0 & 0 & 1 & 0 & 0 & 0 \\
        0 & 0 & 0 & 0 & 0 & 0 & 0 & 0 & 0 \\
        0 & 0 & 0 & 0 & 0 & 0 & 0 & 0 & 1
    \end{bmatrix}.
\end{gathered}\end{equation}
where $\otimes$ denotes the Kronecker product. When $\mathbf F_i = \mathbf I$, the Hessian simplifies to
\begin{equation}\begin{gathered}
    \Hess(\Psi, \mathbf F_i) = \mu \mathbf I + (\lambda + \mu)(2 - \gamma) \mathbf K_1 \\
    - (\lambda + \mu)(1 - \gamma) \mathbf K, \\
    \mathbf K_1 = \begin{bmatrix}
        1 & 0 & 0 & 0 & 1 & 0 & 0 & 0 & 1 \\
        0 & 0 & 0 & 0 & 0 & 0 & 0 & 0 & 0 \\
        0 & 0 & 0 & 0 & 0 & 0 & 0 & 0 & 0 \\
        0 & 0 & 0 & 0 & 0 & 0 & 0 & 0 & 0 \\
        1 & 0 & 0 & 0 & 1 & 0 & 0 & 0 & 1 \\
        0 & 0 & 0 & 0 & 0 & 0 & 0 & 0 & 0 \\
        0 & 0 & 0 & 0 & 0 & 0 & 0 & 0 & 0 \\
        0 & 0 & 0 & 0 & 0 & 0 & 0 & 0 & 0 \\
        1 & 0 & 0 & 0 & 1 & 0 & 0 & 0 & 1
    \end{bmatrix}. \\
    \label{eq:supp_hess_F_simplified}
\end{gathered}\end{equation}
Here $\lambda$ and $\mu$ are Lamé coefficients depending on the sample point $\mathbf X_i$, which we omit for simplicity.
On the other hand, from Eq.~(\ref{eq:supp_F}) we know that
\begin{equation}\begin{gathered}
    \Vector(\mathbf F_i) = \begin{bmatrix}
        \mathbf F_{11} \\ \mathbf F_{12} \\ \mathbf F_{13} \\ \mathbf F_{21} \\ \mathbf F_{22} \\ \mathbf F_{23} \\ \mathbf F_{31} \\ \mathbf F_{32} \\ \mathbf F_{33}
    \end{bmatrix} = \begin{bmatrix}
        \partial_x \boldsymbol\phi(\mathbf X_i)^T \mathbf d^x + 1 \\
        \partial_y \boldsymbol\phi(\mathbf X_i)^T \mathbf d^x \\
        \partial_z \boldsymbol\phi(\mathbf X_i)^T \mathbf d^x \\
        \partial_x \boldsymbol\phi(\mathbf X_i)^T \mathbf d^y \\
        \partial_y \boldsymbol\phi(\mathbf X_i)^T \mathbf d^y + 1 \\
        \partial_z \boldsymbol\phi(\mathbf X_i)^T \mathbf d^y \\
        \partial_x \boldsymbol\phi(\mathbf X_i)^T \mathbf d^z \\
        \partial_y \boldsymbol\phi(\mathbf X_i)^T \mathbf d^z \\
        \partial_z \boldsymbol\phi(\mathbf X_i)^T \mathbf d^z + 1 \\
    \end{bmatrix}.
\end{gathered}\end{equation}
Thus the Jacobian matrix $\mathbf J_i$ is given by
\begin{equation}\begin{gathered}
    \mathbf J_i = \frac{\partial \Vector(\mathbf F_i)}{\partial \mathbf d^x} = \begin{bmatrix}
        \partial_x \boldsymbol\phi(\mathbf X_i)^T \\
        \partial_y \boldsymbol\phi(\mathbf X_i)^T \\
        \partial_z \boldsymbol\phi(\mathbf X_i)^T \\
        0 \\
        0 \\
        0 \\
        0 \\
        0 \\
        0 \\
    \end{bmatrix} \in \mathbb R^{9 \times K}.
    \label{eq:supp_J}
\end{gathered}\end{equation}
Substituting Eq.~(\ref{eq:supp_hess_F_simplified}) and Eq.~(\ref{eq:supp_J}) into Eq.~(\ref{eq:supp_hess_dx}), we get
\begin{equation}\begin{gathered}
    \Hess(\Psi(\mathbf F_i), \mathbf d^x) = \mu \sum_{s=x,y,z} \partial_s \boldsymbol\phi(\mathbf X_i) \partial_s \boldsymbol\phi(\mathbf X_i)^T \\
    + (\lambda + \mu)(2 - \gamma) \partial_x \boldsymbol\phi(\mathbf X_i) \partial_x \boldsymbol\phi(\mathbf X_i)^T \\
    + (\lambda + \mu)(\gamma - 1) \partial_x \boldsymbol\phi(\mathbf X_i) \partial_x \boldsymbol\phi(\mathbf X_i)^T \\
\end{gathered}\end{equation}
With $\gamma = 1 + \mu/(\lambda + \mu)$, we have
\begin{equation}\begin{gathered}
    \Hess(\Psi(\mathbf F_i), \mathbf d^x) = (\lambda + 2\mu) \partial_x \boldsymbol\phi(\mathbf X_i) \partial_x \boldsymbol\phi(\mathbf X_i)^T \\
    + \mu \partial_y \boldsymbol\phi(\mathbf X_i) \partial_y \boldsymbol\phi(\mathbf X_i)^T + \mu \partial_z \boldsymbol\phi(\mathbf X_i) \partial_z \boldsymbol\phi(\mathbf X_i)^T.
\end{gathered}\end{equation}
Similarly,
\begin{equation}\begin{gathered}
    \Hess(\Psi(\mathbf F_i), \mathbf d^y) = (\lambda + 2\mu) \partial_y \boldsymbol\phi(\mathbf X_i) \partial_y \boldsymbol\phi(\mathbf X_i)^T \\
    + \mu \partial_x \boldsymbol\phi(\mathbf X_i) \partial_x \boldsymbol\phi(\mathbf X_i)^T + \mu \partial_z \boldsymbol\phi(\mathbf X_i) \partial_z \boldsymbol\phi(\mathbf X_i)^T, \\
    \Hess(\Psi(\mathbf F_i), \mathbf d^z) = (\lambda + 2\mu) \partial_z \boldsymbol\phi(\mathbf X_i) \partial_z \boldsymbol\phi(\mathbf X_i)^T \\
    + \mu \partial_x \boldsymbol\phi(\mathbf X_i) \partial_x \boldsymbol\phi(\mathbf X_i)^T + \mu \partial_y \boldsymbol\phi(\mathbf X_i) \partial_y \boldsymbol\phi(\mathbf X_i)^T.
\end{gathered}\end{equation}
In conclusion,
\begin{equation}\begin{gathered}
    \mathbf H_w = \mathbf H_{xx} + \mathbf H_{yy} + \mathbf H_{zz} \\
    = \sum_{i} \sum_{s=x,y,z} v_i \Hess(\Psi(\mathbf F_i), \mathbf d^s) \\
    = \sum_{i} \sum_{s=x,y,z} v_i (\lambda + 4\mu)\partial_s \boldsymbol\phi(\mathbf X_i) \partial_s \boldsymbol\phi(\mathbf X_i)^T \\
    = \sum_{i} v_i (\lambda + 4\mu) \nabla \boldsymbol\phi(\mathbf X_i) \nabla \boldsymbol\phi(\mathbf X_i)^T \\
    \approx \int_\Omega (\lambda(\mathbf X) + 4\mu(\mathbf X)) \nabla \boldsymbol\phi(\mathbf X) \nabla \boldsymbol\phi(\mathbf X)^T d\mathbf X,
\end{gathered}\end{equation}
The $(i, j)$-th element of $\mathbf H_w$ is
\begin{equation}\begin{gathered}
    (\mathbf H_w)_{ij} = \int_\Omega (\lambda(\mathbf X) + 4\mu(\mathbf X)) \nabla \phi_i(\mathbf X)^T \nabla \phi_j(\mathbf X) d\mathbf X.
\end{gathered}\end{equation}
This completes the proof of Proposition~\ref{prop:hessian}.

\begin{table}[t]
  \centering
  \begin{tabular}{l l c c}
  \toprule
Test & $m$ & Simplicits & Ours \\
  \midrule
\multirow{4}{*}{Bend} & 6  & 4.54e-06 & 6.67e-07  \\ 
                      & 9  & 1.19e-06 & 4.33e-07 \\ 
                      & 16 & 5.34e-07 & 1.40e-07 \\ 
                      & 32 & 1.93e-07 & \textbf{5.60e-08} \\ 
  \midrule
\multirow{4}{*}{Twist} & 6 & 9.41e-05 & 2.20e-05 \\ 
& 9 & 1.52e-05 & 5.94e-06 \\ 
& 16 & 5.19e-06 & 5.83e-07 \\ 
& 32 & 7.56e-07 & \textbf{3.29e-07}  \\ 
  \bottomrule
  \end{tabular}
  \caption{Comparison of basis fitting residual for reduced order methods (Simplicits and ours) for the standard beam test.}
  \label{table:residual_beam}
\end{table}

\begin{table*}[t]
  \centering
  \begin{tabular}{c c c c c c c c}
  \toprule
\multirow{2}{*}{Method} & \multicolumn{2}{c}{Fix Side} & \multicolumn{2}{c}{Pull Farthest} & \multicolumn{2}{c}{Pull Boundary}\\
& MSE & Max & MSE & Max & MSE & Max \\
  \midrule 
Simplicits & 1.64e-05 & 3.60e-05 & 1.15e-04 & 2.86e-04 & 4.62e-05 & 9.81e-05 \\
Ours & \textbf{4.55e-06} & \textbf{1.00e-05} & \textbf{3.52e-05} & \textbf{8.84e-05} & \textbf{1.30e-05} & \textbf{2.63e-05} \\
Improvement & 72.3\% & 72.1\% & 69.5\% & 69.1\% & 71.9\% & 73.2\% \\               
  \midrule
Simplicits & 5.71e-12 & 6.77e-12 & 4.74e-05 & 1.35e-04 & 7.31e-05 & 2.25e-04 \\
Ours & \textbf{1.90e-12} & \textbf{2.51e-12} & \textbf{7.51e-06} & \textbf{2.18e-05} & \textbf{9.98e-06} & \textbf{2.92e-05} \\
Improvement & 66.8\% & 62.9\% & 84.2\% & 83.9\% & 86.3\% & 87.0\% \\
  \bottomrule
  \end{tabular}
  \caption{Basis fitting residual error on the Thingi10K and Simready Datasets. We compute the least square fitting of the FEM simulation results using the predicted skinning weights from Simplicits and our method, and report the fitting residual to quantify the capability of those skinning weights to express the full-order FEM deformation. Our results show consistent improvement over the Simplicits baseline.}
  \label{table:residual_thingi10k_simready}
\end{table*}

\section{Additional Evaluation and Analysis}

\subsection{Basis Fitting Residual}

For reduced-order methods including Simplicits and ours, we also perform a least square fitting of the FEM simulation results using the predicted skinning weights, and report the fitting residual on the standard beam test in Table \ref{table:residual_beam} and on the Thingi10K/Simready datasets in Table \ref{table:residual_thingi10k_simready}. 

Concretely, given $N$ rest-post vertices of the FEM mesh $\{\mathbf X_i\}_{i=1}^N$ and the corresponding deformed positions $\mathbf x_{i,t}^{\text{FEM}}$ at time $t$ from the full-order FEM simulation, we fit the reduced-order DoFs $\mathbf z$ (the $m$ affine transformations in Eq.~(\ref{eq:method_skinning})) by minimizing the squared displacement error:
\begin{equation}
\mathbf z_t^* = \argmin_{\mathbf z} \frac{1}{N} \sum_{i=1}^N \left\| \Phi(\mathbf X_i, \mathbf z) - \mathbf x_{i,t}^{\text{FEM}} \right\|^2,
\label{eq:supp_fitting}
\end{equation}
where $\Phi(\mathbf X, \mathbf z)$ is the skinning deformation map in Eq.~(\ref{eq:method_skinning}). Notice that in Simplicits and our method, the deformation map $\Phi$ is linear with respect to $\mathbf z$, so Eq.~(\ref{eq:supp_fitting}) is a linear least square problem and can be solved analytically. With the optimal fitted deformation $\mathbf z_t^*$, we report the average residual
$$ \frac1{NT} \sum_{i=1}^N \sum_{t=1}^T \left\| \Phi(\mathbf X_i, \mathbf z_t^*) - \mathbf x_{i,t}^{\text{FEM}} \right\| $$
over the $N$ vertices and $T$ time steps, with the same normalization by bounding box size as described in Sec.~\ref{sec:evaluation}.

Compared with the mean squared error measured on the simulated deformation in Sec.~\ref{sec:evaluation}, this metric factors out the difference in dynamic behavior accumulated during the simulation, and only evaluates the expressiveness of the predicted skinning weights. The results in Table \ref{table:residual_beam} and \ref{table:residual_thingi10k_simready} demonstrate that our skinning weights, or simulation basis, can better explain the FEM simulation results than the Simplicits baseline.

\begin{figure}[t]
    \centering
    \includegraphics[width=\linewidth]{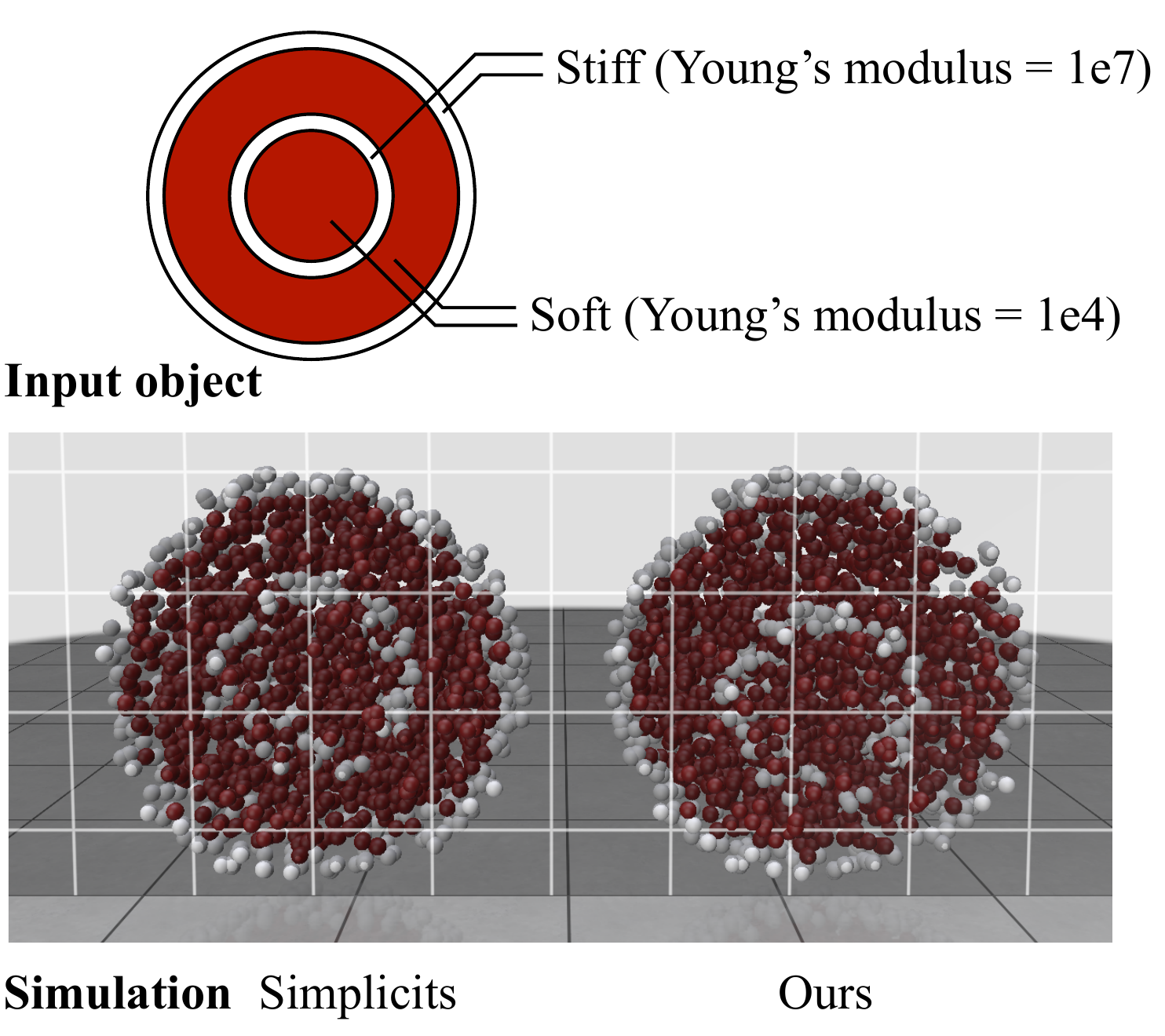}
    \caption{
        We compare our method with Simplicits on a heterogeneous sphere consisting of four layers of stiff-soft-stiff-soft materials. The hard layers are shown in white and the soft layers in red. We visualize the internal deformation of the spheres using a slice plane that passes through the sphere centers. 
    }
    \label{fig:result_multi_layer}
\end{figure}

\subsection{Multi-Layer Material Simulation}

In Fig.~\ref{fig:result_multi_layer}, we compare our method with Simplicits on a heterogeneous sphere consisting of four layers of stiff-soft-stiff-soft materials. During simulation, the sphere is dropped to the ground. This case is challenging because it requires the global simulation basis to express the distinct dynamic patterns across different stiffness regions. Our method captures realistic and distinct deformation behaviors across the varying layers. In contrast, the Simplicits simulation only produces stiff, global deformation. This result is best viewed in the supplementary video.

\begin{figure}[t]
    \centering
    \includegraphics[width=\linewidth]{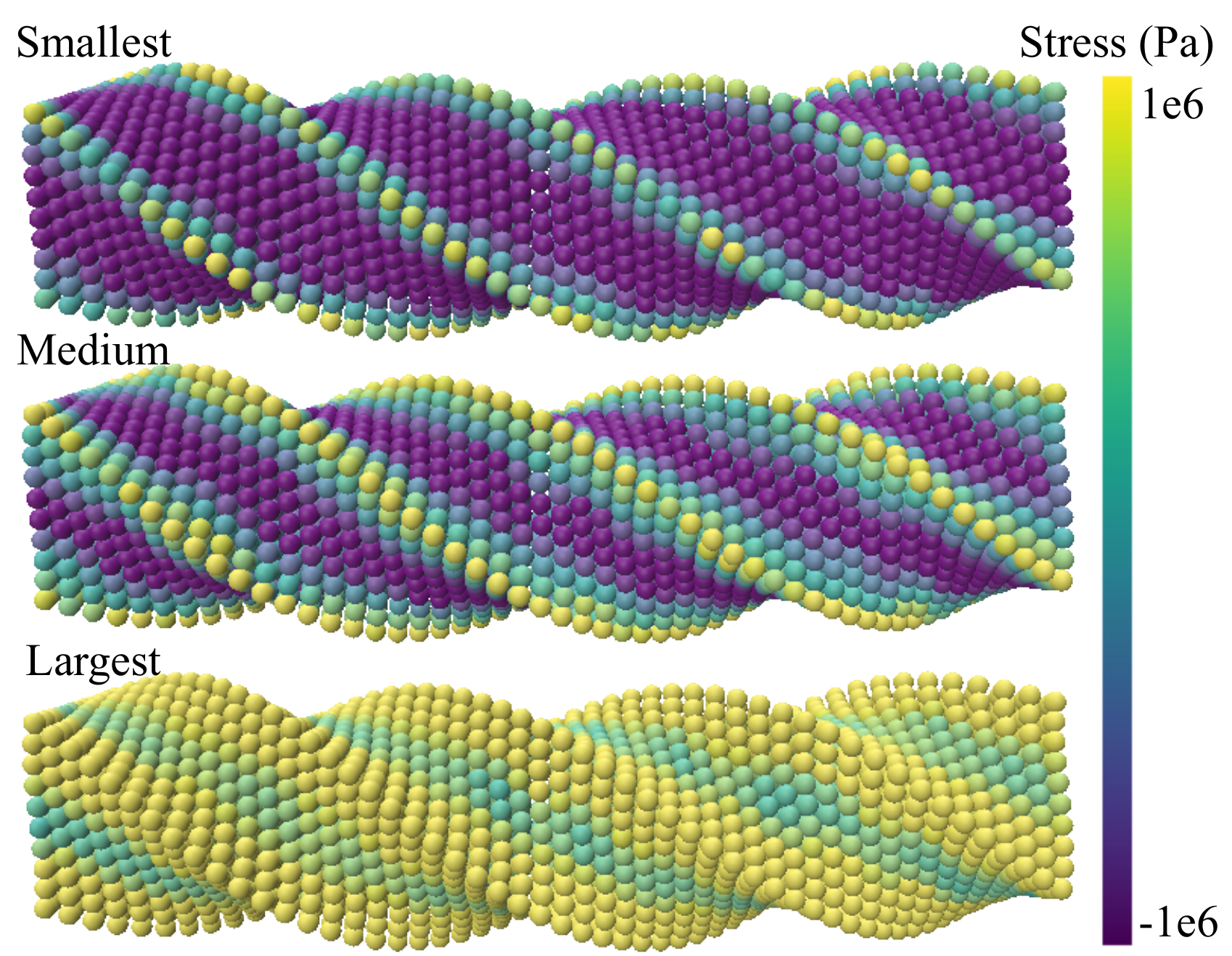}
    \caption{
        The Cauchy stress distribution on the twisted beam test. We plot the three principal stresses components of the Cauchy stress tensor capped within the range of $[-\num{1e6}, \num{1e6}]$ Pa.
    }
    \label{fig:result_stress}
\end{figure}

\subsection{Stress Visualization}

In Fig.~\ref{fig:result_stress}, we visualize the stress distribution on the beam in the twist test. Concretely, we convert the first Piola-Kirchhoff stress tensor $\mathbf P = \frac{\partial \Psi}{\partial \mathbf F}$ defined in the reference configuration into Cauchy stress tensor in the deformed configuration using the relation
$$ \boldsymbol \sigma = \frac1{\det \mathbf F} \mathbf P \mathbf F^T, $$
which is a symmetric $3 \times 3$ tensor at each point over the deformed beam. We plot the three principal stress fields (the eigenvalues of $\boldsymbol \sigma$) from lowest to highest in Fig.~\ref{fig:result_stress}. This result allows us to visualize the distribution of stress within the beam, and identify the regions of high stress concentration. We hope this will inspire more results on mechanical analysis on top of reduced-order simulation in the future.

\end{document}